\documentclass[letterpaper, 10 pt, conference]{ieeeconf}  
\usepackage{comment}       
\usepackage{cite}
\usepackage{amsmath,amssymb,amsfonts}
\usepackage{algorithmic}
\usepackage{graphicx}
\usepackage{textcomp}
\usepackage{xcolor}
\usepackage{makecell}
\usepackage{url} 
\usepackage[hidelinks]{hyperref}

\IEEEoverridecommandlockouts                              

\title{\LARGE \bf
BCI-Based Assessment of Ocular Response Time Using Dynamic Time Warping Leveraging an RDWT-Driven Deep Neural Framework
}

\author{S. Sarkar$^{1}$, 
~S. S. Gandavarapu$^{2}$, 
~Jeff Feng$^{3}$,
~Saurabh Prasad$^{4}$, 
~Reza Khanbabaie$^{5}$ and  
~Jose L. Contreras-Vidal$^{6}$
\thanks{*This work was supported by NSF IUCRC BRAIN award \# 2137255 and NSF REU site award \#2150415.}
\thanks{$^{1}$Shantanu Sarkar is a doctoral candidate, Dept. of ECE, Univ. of Houston, Houston, TX, USA, {\tt\small shantanu75@gmail.com}}%
\thanks{$^{2}$Sai S. Gandavarapu is a graduate student, Dept. of Data Science, Univ. of Houston, Houston, TX, USA, {\tt\small sgandava@cougarnet.uh.edu}}%
\thanks{$^{3}$Jeff Feng is with Faculty of Industrial Design, Univ. of Houston, Houston, TX, USA, {\tt\small ffeng@central.uh.edu}}%
\thanks{$^{4}$Saurabh Prasad is with Faculty of ECE, Univ. of Houston, Houston, TX, USA, {\tt\small sprasad2@central.uh.edu}}%
\thanks{$^{5}$Reza Khanbabaie is a Neuroscientist with Cognixion Inc., Toronto, Ontario, Canada, {\tt\small reza.khanbabaie@cognixion.com}}%
\thanks{$^{6}$Jose L. Contreras-Vidal is with Faculty of ECE, Univ. of Houston, Houston, TX, USA, {\tt\small jlcontreras-vidal@uh.edu}}%
}

\begin{document}

\maketitle
\thispagestyle{empty}
\pagestyle{empty}

\begin{abstract}
Mild traumatic brain injury (mTBI) is a prevalent condition that remains difficult to diagnose in its early stages. Oculomotor dysfunction is a well-established marker of mTBI, motivating the development of portable tools that capture both eye-movement behavior and underlying neurophysiology. In this work, we present an initial framework that integrates electroencephalogram (EEG) with augmented-reality (AR)-based Vestibular/Ocular Motor Screening (VOMS) tasks to estimate subject-specific ocular response times. Pre-processed EEG signals, obtained through band-pass filtering and average referencing, are analyzed using a Redundant Discrete Wavelet Transform (RDWT)–driven deep neural framework. The RDWT coefficients are subjected to trainable zero-phase convolutional filtering and reconstructed into the time domain via inverse RDWT, followed by channel-wise temporal and spatial filtering using 2D convolution layers and convolutional–LSTM-based decoding. An ablation study demonstrates that wavelet-domain filtering serves as an effective denoising strategy, improving prediction performance. Sliding-window predictions were validated using Pearson correlation ($> 0.5$), and Dynamic Time Warping (DTW) was subsequently used to estimate ocular response times. DTW-derived metrics revealed VOM-task-dependent differences in ocular response time between participants, characterized using Mann–Whitney U tests. Cross-correlation analysis further revealed task-dependent temporal behaviors: pursuit tasks exhibited reactive tracking, whereas saccades showed anticipatory responses. Overall, the results highlight pursuit tasks as particularly informative for distinguishing timing differences and demonstrate the potential of RDWT-based EEG features combined with DTW metrics for multimodal mTBI assessment.
\end{abstract}

\section{INTRODUCTION}
\begin{figure}[t!]
\centering
\includegraphics[width=0.95\linewidth]{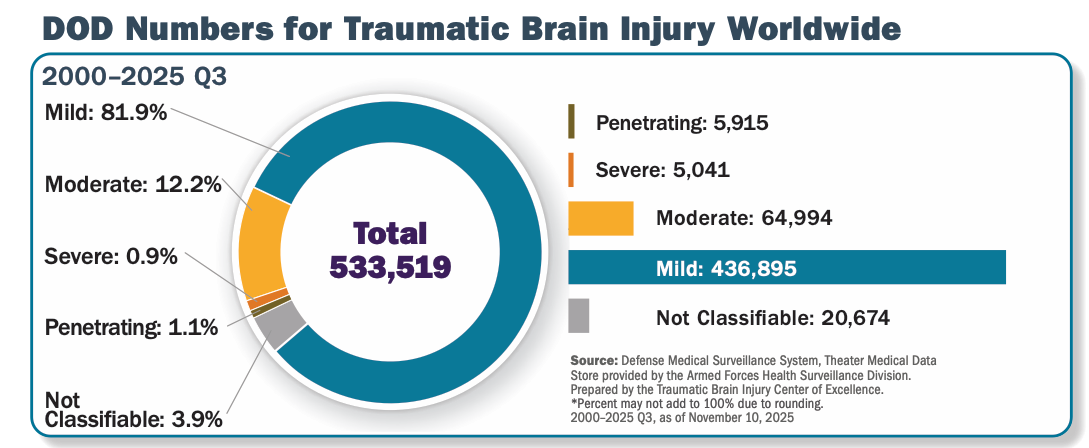}
\caption{Department of Defense (DoD) worldwide traumatic brain injury cases from 2000 to Q3 2025, reported by the Defense Health Agency \cite{DoD_TBI_No}.}
\label{fig1}
\vspace{-10pt}
\end{figure}

Traumatic brain injury (TBI) refers to a disruption of normal brain function caused by a bump, blow, jolt, or penetrating head injury \cite{CDC2025}. Military service members (SMs) face elevated risks of TBI from falls, car accidents, strikes, and blast exposures in combat or training. In the US and worldwide, millions of concussion cases occur among civilians. Based on symptoms and severity, TBI is classified into three levels: mild, moderate, and severe. Mild TBI (mTBI) symptoms include headache, dizziness, nausea, cognitive fog, and memory loss. According to the Defense and Veterans Brain Injury Center (DVBIC), more than 528,450 TBIs among U.S. SMs worldwide have been reported between 2000 and 2025, with over 81.8\% classified as mild (mTBI) (Fig.~\ref{fig1}) \cite{DoD_TBI_No}. Despite its prevalence and extensive diagnostic research, mTBI remains poorly understood and difficult to diagnose \cite{McKee2014}. Computerized cognitive tests such as ImPACT \cite{ImPACT},  symptom checklists such as the Post-Concussion Symptom Scale (PCSS) \cite{PCSS}, and clinical assessments with some scoring system like the Balance Error Scoring System (BESS) \cite{BESS} are commonly used for evaluating concussions.
\\
While these tools are widely used in both clinical and athletic settings, concerns persist regarding their reliability, particularly in the absence of baseline metrics, their test-retest reliability over time, and limitations imposed by other psychological and physical conditions \cite{Farnsworth2017, Resch2013, Mason2020}. However, studies using ImPACT have shown moderately reliable metrics for visual-motor speed and reaction time \cite{Resch2013, Mason2020}. Neurophysiological methods such as electroencephalogram (EEG) and electrooculography (EOG) capture brain dynamics and ocular movements, yet their clinical application is hindered by low signal-to-noise ratios (SNR) and artifacts \cite{Uriguen2015,Craik2023}. Given the time-sensitive nature of TBI, rapid, field-applicable tools are essential. However, no effective method or product is available to address this gap \cite{Kutcher2014}. Practicing effective tracking of TBI recovery over time remains challenging, as no existing device is suited for this task \cite{Munia2017, Yue2020}. The study demonstrated that the Vestibular/Ocular Motor Screening (VOMS) assessment is a reliable tool for identifying patients with concussion \cite{Mucha2014}.
\\
Based on these insights, we propose `\textbf{SynTec}', a lightweight, wearable multimodal approach that integrates an augmented-reality (AR) display, eye tracking, and EEG/EOG into a single portable platform for neuro-ocular response time assessment, aimed at supporting mTBI evaluation and longitudinal recovery monitoring. Preliminary prototype testing revealed consistent neurophysiological and oculomotor trends across subjects, suggesting the feasibility of a rapid, field-deployable approach to brain assessment for time-sensitive conditions such as mTBI and concussion. \textbf{SynTec} is intended to support field-based evaluation of visual–motor reaction time and associated neurophysiological signals. Our AI-driven signal processing and boosting techniques will provide clear, real-time analysis of multimodal data, enabling medics and patients make rapid, confident calls about mTBI. Beyond initial diagnosis, \textbf{SynTec} will establish personalized baseline profiles and track brain recovery in real time over weeks to months, offering clearer insight into healing progress. With its modular and flexible design, \textbf{SynTec} has the potential to adapt to other conditions involving eye movement and coordination deficits, such as autism spectrum disorder and Parkinson's disease, where eye-tracking and reaction tests are established tools for diagnosis \cite{ Toda1993, vanderGeest2001, Wylie2009,Zhang2016,Xie2024}.
\\
For the proof of concept (PoC), we developed an initial framework and evaluated it using two control subjects across four VOMS tasks encompassing both saccades and pursuits in the horizontal and vertical directions. Our results showed that pursuit paradigms yielded the most distinguishable ocular response times between participants when leveraging a Redundant Discrete Wavelet Transform (RDWT) driven deep neural framework for EEG feature extraction, together with Dynamic Time Warping (DTW) based latency metrics.
\section{Method}
\subsection{Experimental Design}
The study included two control participants who met the inclusion criteria of being between 18 and 65 years of age, with no known history of mild traumatic brain injury (mTBI), no confounding neurological conditions such as epilepsy, and no visual impairments. Demographic information for both participants is summarized in Table~\ref{tab:Table1}. Data collection was conducted using the Axon-R device (Cognixion Inc., Santa Barbara, CA), a brain–computer interface platform that integrates augmented reality (AR) with six-channel EEG monitoring.
\begin{table}[ht!]
\centering
\renewcommand{\arraystretch}{1.3} 
\caption{Participant demographics (gender and age).}
\label{tab:Table1}
\begin{tabular}{ccc}
\hline
\noalign{\vskip 0.5ex}
\textbf{Participant} & \textbf{Gender} & \textbf{Age (years)} \\
\noalign{\vskip 0.5ex}
\hline
S1 & M & 55 \\
S2 & M & 23 \\
\hline
\vspace{-10pt}
\end{tabular}
\end{table}
\begin{figure}[!t]
\centering
\includegraphics[width=0.92\linewidth]{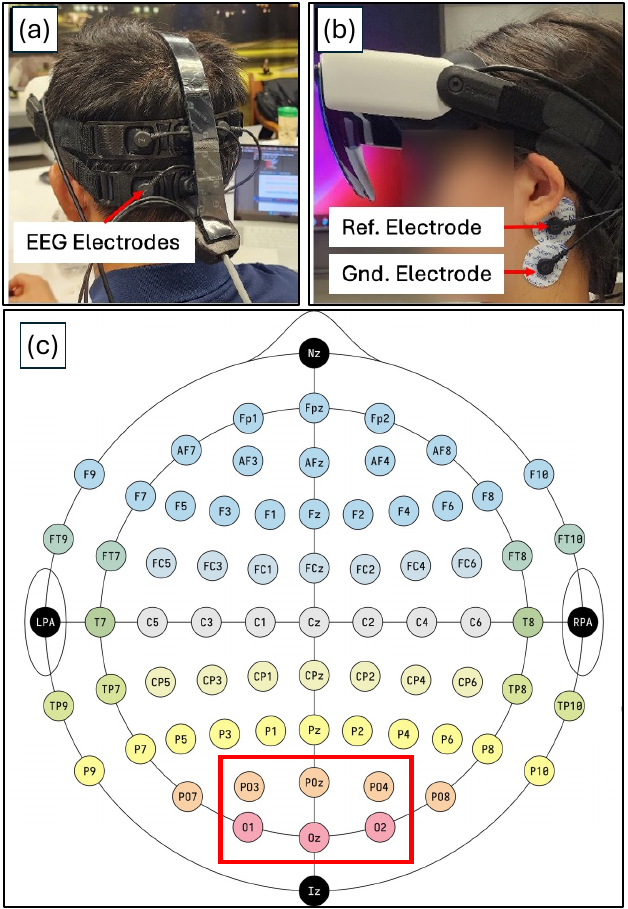}
\caption{EEG acquisition setup and electrode configuration. (a) Participant wearing an Axon-R headset showing the placement of EEG electrodes. (b) Participant wearing Axon-R headset showing the placement of Ref. and Gnd. electrodes. (c) Highlighted EEG electrode positions (red box) used in the experiment, focused on occipital and parietal-occipital regions.}
\label{fig2}
\vspace{-1pt}
\end{figure}
\begin{figure}[!b]
\vspace{-5pt}
\centering
\includegraphics[width=\linewidth]{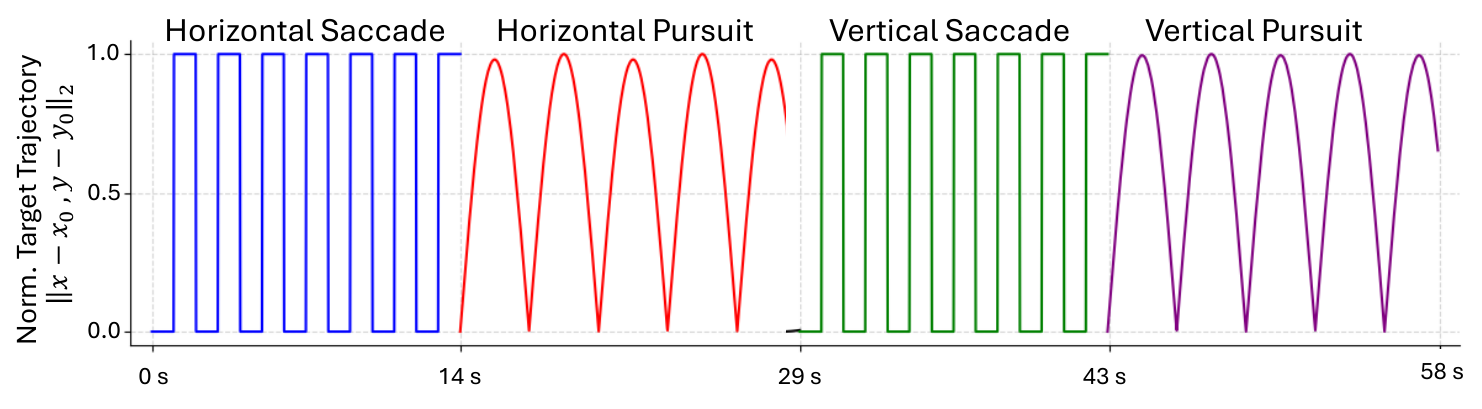}
\vspace{-5pt}
\caption{Normalized target trajectory computed as the L2 distance between the centers of the AR patch and AR window, resampled at 100 Hz.}
\label{fig3}
\end{figure}
\\Participants completed four Vestibular/Ocular Motor Screening (VOMS) tasks, described in subsequent subsections. Each task required the participant to visually track a target (circular white patch) presented within the AR display. EEG signals were simultaneously recorded from six electrode positions (O1, Oz, O2, PO3, POz, and  PO4) as shown in Fig.~\ref{fig2}. At the beginning of every session, electrode impedance was manually checked to ensure values remained below 20 $k\Omega$. To achieve electrode impedances below 20 $k\Omega$, electrode gel - Spectra 360 (Parker Laboratories Inc., Fairfield, NJ) and electrode solution, and Skin prep - Signa Spray (Parker Laboratories  Inc., Fairfield, NJ) were applied to reduce contact impedance.
\\
Each session consisted of five trials. Each trial lasted 60 seconds and included all four VOMS tasks. The initial 2 seconds were used to instruct the participant to relax and perform some jaw clenching, allowing visual validation of the quality of the EEG signals. The remaining 58 seconds were dedicated to the VOMS tasks. Data were collected across ten sessions per participant. 
\subsubsection{Vestibular/Ocular Motor Screening (VOMS)}
mTBI is often characterized by abnormal eye-movement behavior, and prior research has documented visual and oculomotor deficits in individuals with mTBI across pediatric and adult populations. In a prior study, Mucha et al. assessed 64 patients with concussion using Vestibular/Ocular Motor Screening (VOMS) assessment, encompassing five domains: smooth pursuit, horizontal and vertical saccades, near point of convergence, horizontal vestibulo-ocular reflex, and visual motion sensitivity \cite{Mucha2014}. Motivated by this framework, in this initial study, we implemented four VOMS tasks—horizontal and vertical smooth pursuits and saccades. These were selected because a smoothly rendered, controlled moving visual target can be presented and measured reliably in AR without additional calibration.
\\
\textbf{Horizontal Saccade:} Each horizontal saccade trial lasted 14 seconds. A white circular patch appeared on the AR display and alternated between the left and right extremes at 1‑second intervals, requiring participants to shift their gaze between the two positions rapidly.
\\
\textbf{Horizontal Pursuit:} For the horizontal pursuit task, the white circular patch moved smoothly from the center to the right extreme, then returned to the center, and repeated the same motion toward the left extreme. A single excursion to one extreme and back required approximately 3.16 seconds, and each pursuit trial lasted a total of 15 seconds.
\\
\textbf{Vertical Saccade:} Vertical saccade trials also lasted 14 seconds and followed the same structure as horizontal saccades. The white circular patch alternated between the top and bottom extremes at 1‑second intervals.
\\
\textbf{Vertical Pursuit:} The vertical pursuit task followed an analogous pattern, with the target moving smoothly from the center to the top extreme and back, then to the bottom extreme and back. As with the horizontal pursuit, one full excursion required approximately 3.16 seconds and each trial lasted 15 seconds.
\\
During the horizontal and vertical saccade tasks, participants were instructed to shift their gaze to follow the target's abrupt positional changes. During the horizontal and vertical pursuit tasks, participants were instructed to continuously follow the smoothly moving target within the AR environment. The AR system updated the VOMS stimuli at 60 Hz. For analysis, the target trajectory was represented as the L2 distance between the centers of the circular white patch and the AR window, computed from their `\textit{x}' and `\textit{y}' coordinates and resampled at 100 Hz, as illustrated in Fig.~\ref{fig3}.
\subsubsection{Electroencephalogram (EEG)}
During each trial, EEG signals were recorded from six Ag/AgCl electrodes mounted on the Axon-R device. Three electrodes (O1, Oz, O2) were positioned over the occipital region, and three electrodes (PO3, POz, PO4) were positioned over the parieto-occipital region, as shown in Fig.~\ref{fig2}.  At the beginning of each session, electrode channel impedances were manually verified to be below 20 $k\Omega$. EEG data were acquired using the Axon-R device, which samples at 250 Hz (Axon-R sampling rate).
\begin{figure*}[!t]
\centering
\includegraphics[width=\textwidth]{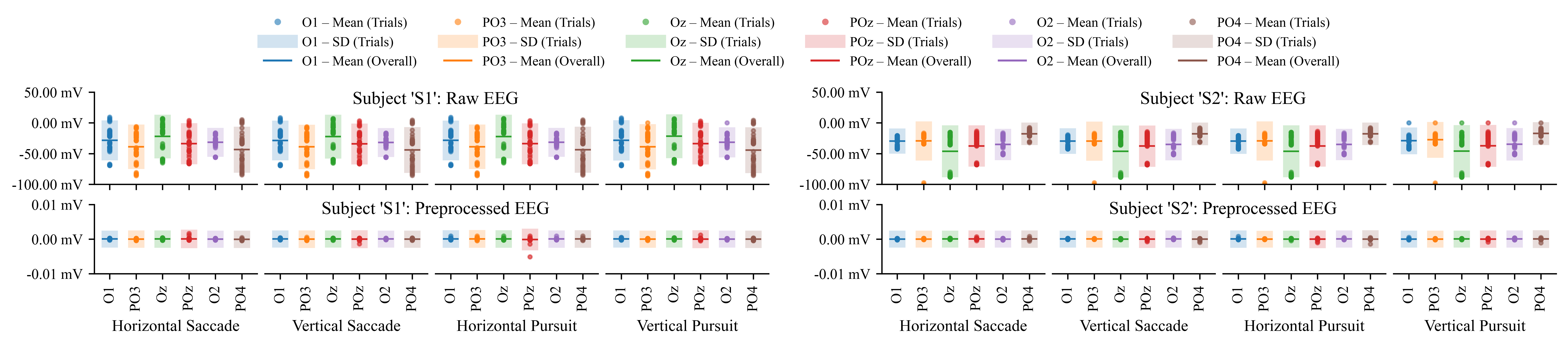}
\caption{Signal quality of raw and preprocessed EEG data was verified using trial‑wise mean and standard deviation plots across channels and participants.}
\label{fig4}
\end{figure*}
\subsection{EEG Preprocessing}
The EEG signals were down-sampled to 100 Hz, limiting the frequency component range within 50 Hz, and aligned with the target trajectory, represented by the L2 distance computed from the `\textit{x}' and `\textit{y}' coordinates of the center of the circular white patch. Additionally, to filter out high-frequency artifacts, we used a zero‑phase, 4th‑order Butterworth band‑pass filter (BPF) with cutoff frequencies of 0.5 Hz and 30 Hz, followed by average re‑referencing. We intended to exclude the $\gamma$ band, as frequencies above 25 Hz are highly susceptible to motion-related artifacts \cite{Pope2022}. To verify the signal quality of preprocessed EEG data, the mean and standard deviation for each trial were plotted, as shown in Fig.~\ref{fig4}. 
\\
A sliding window approach was applied using a 2.56 s window and a 200 ms stride, so we considered only the first 13.76 s of data across all VOMSs, yielding 57 sliding windows per VOMS per trial. For S1, the EEG recording ended slightly early in \#18, leaving only 13.69 sec of vertical pursuit data, while for S2, early termination in \#5 resulted in just 9.87 s; as a result, the vertical pursuit data from S1 in \#18 and S2 in \#5 were omitted from the analysis. Consequently, for each subject across 50 trials, we extracted 2,793 windows for the vertical pursuit task and 2,850 windows for each of the remaining VOMSs. 
\begin{figure}[!hb]
\vspace{-10pt}
\centering
\includegraphics[width=0.70\linewidth]{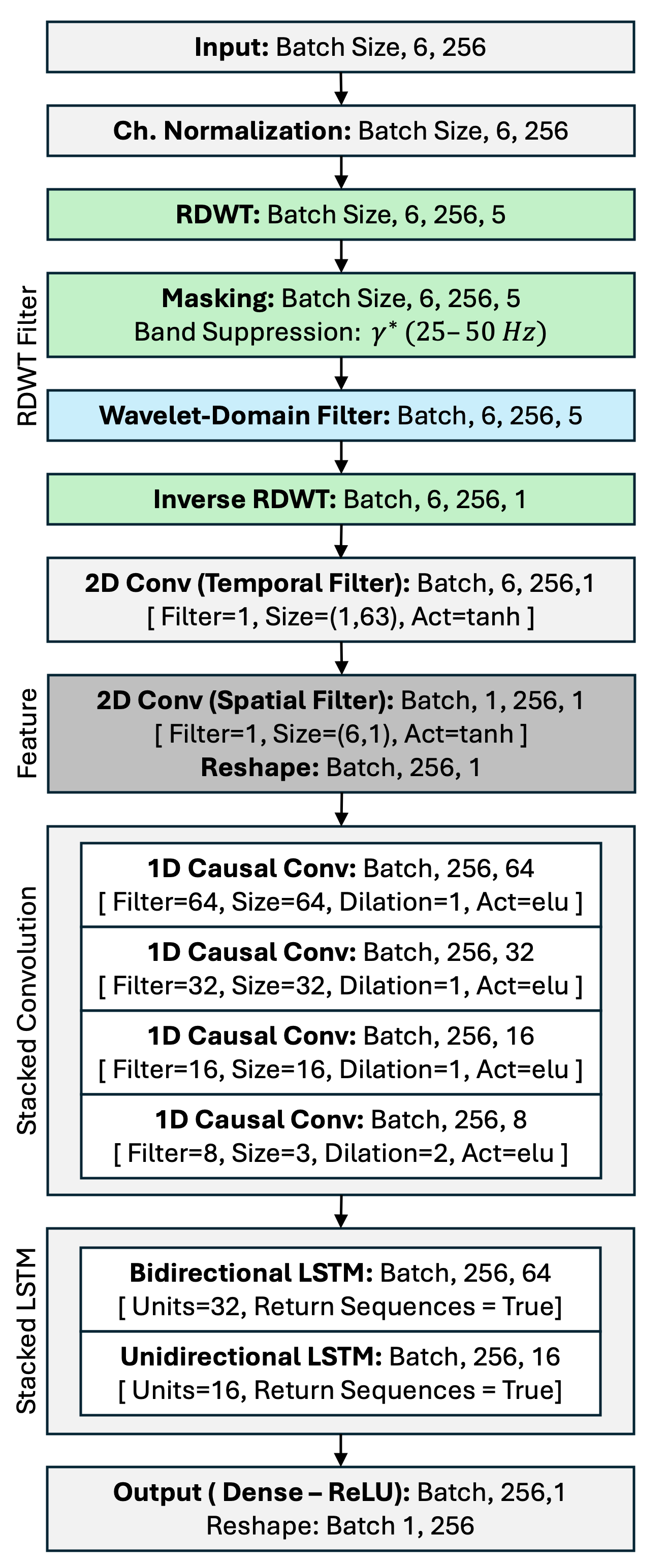}
\caption{Model architecture illustrating RDWT‑inspired band‑specific filtering using Symlet‑2 MW, temporal and spatial filtering via convolution layers, and stacked Conv–LSTM and Dense layers for target trajectory prediction.}%
\label{fig5}
\end{figure}
\begin{figure}[t]
\centering
\includegraphics[width=\linewidth]{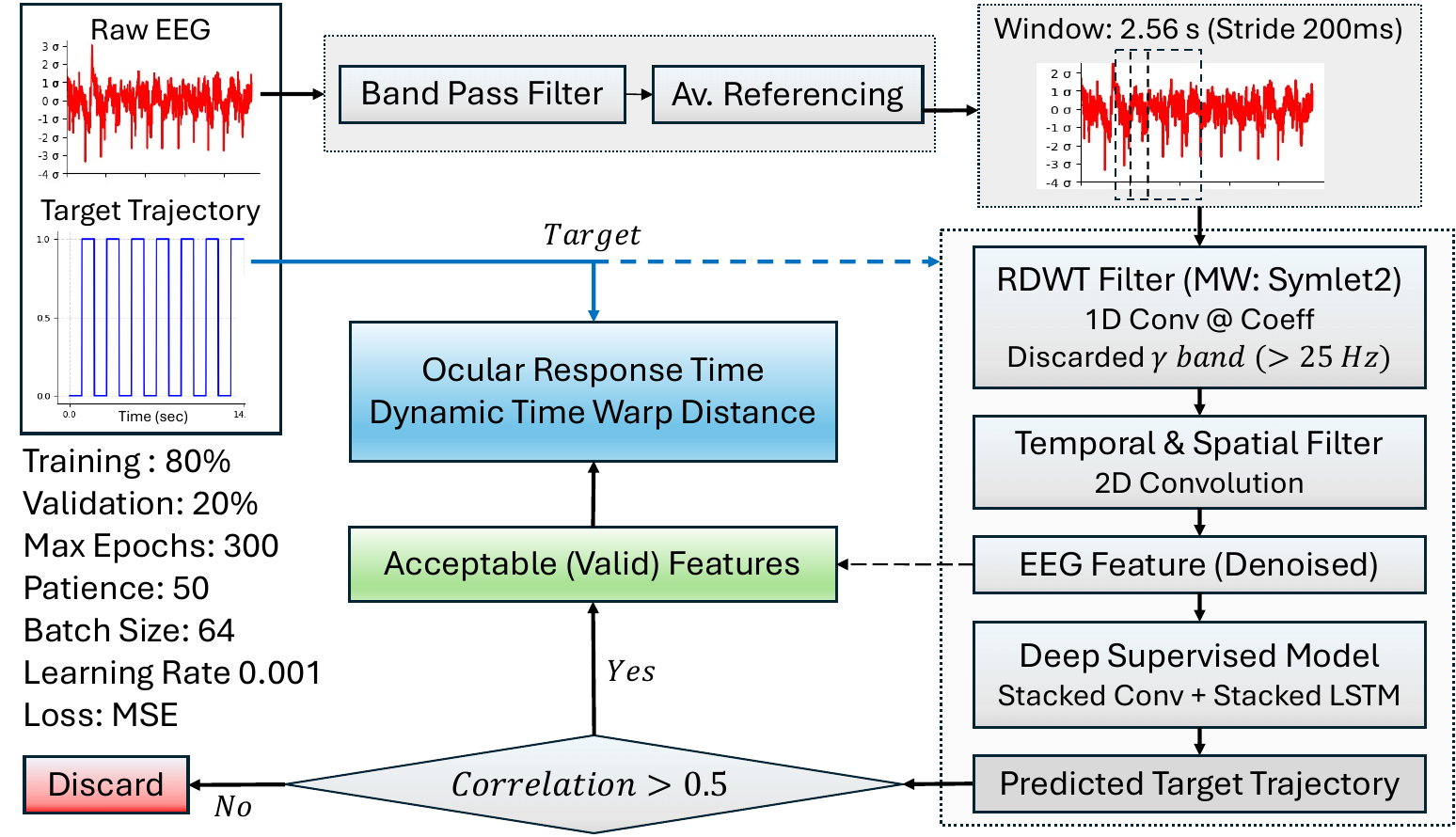}
\caption{Block diagram of the analysis pipeline, including preprocessing, feature extraction, supervised modeling, and correlation-based trial validation.}
\label{fig6}
\vspace{-20pt}
\end{figure}
\subsection{Analysis Methodology}
The analysis employed a deep learning framework incorporating a Redundant Discrete Wavelet Transform (RDWT)- inspired filtering stage using the Symlet-2 Mother Wavelet (MW). The choice of the Symlet-2 MW was motivated by our previous comparative analysis, which confirmed that Symlet-2 yields the highest correlation between reconstructed and original EEG signals across the $\delta^*$ (0–3.125 Hz), $\theta^*$ (3.125–6.25 Hz), $\alpha^*$ (6.25–12.5 Hz), and $\beta^*$ (12.5–25 Hz) bands, considering both magnitude and phase spectral characteristics \cite{Sarkar2025}. We introduced a custom layer for 0-phase-shift band-specific filtering of wavelet coefficients using two 1D Convolution layers, followed by an inverse-RDWT for reconstructing the denoised EEG channels. The reconstructed signals are then processed by two consecutive 2D convolution layers: the first performs channel-wise temporal filtering, while the second applies spatial filtering across channels, resulting in a compact 1D EEG feature representation. Down the line, the Model used stacked 1D convolutional neural network,  stacked LSTM, and Dense layers to predict the target trajectory window.
\\
We used 2.56 s sliding windows of EEG signals, along with the target trajectory represented by L2 distance, to train the deep learning model. Using the trained Model, we predicted the target trajectory window, and the windows showing a correlation coefficient $>$0.5 were further considered for ocular response time analysis. The architecture of the deep learning framework incorporating wavelet-domain filtering is illustrated in Fig.~\ref{fig5} and discussed in Section~\ref{sec:rdwt_model} below, and Fig.~\ref{fig6} presents the full analysis pipeline.
\begin{table}[hb]
\vspace{-5pt}
\centering
\caption{Kernel sizes and corresponding effective frequency scales for wavelet-domain filtering.}
\label{tab:Table2}
\renewcommand{\arraystretch}{1.3} 
\begin{tabular}{c c c}
\hline
\noalign{\vskip 0.8ex}
\makecell{\textbf{Sub-band} \\[0.4ex] \textbf{(Freq. Range)}} &
\makecell{\textbf{Kernel Size} \\[0.4ex] \textbf{(K)}} &
\makecell{\textbf{Effective Freq.} \\[0.4ex] 
\small{$f_{\mathrm{eff}} \approx 100/K\;\text{Hz}$}} \\
\noalign{\vskip 0.8ex}
\hline
$\delta^{*}$ (0--3.125 Hz)    & 63 & 1.6  \\
$\theta^{*}$ (3.125--6.25 Hz) & 15 & 6.7  \\
$\alpha^{*}$ (6.25--12.5 Hz)  & 7  & 14.3 \\
$\beta^{*}$ (12.5--25 Hz)     & 3  & 33.3 \\
\hline
\end{tabular}
\end{table}
\subsubsection{RDWT--Integrated Deep Learning Model}
\label{sec:rdwt_model}
The Model's input layer is designed to accept 6-channels of 256 samples (2.56 s). The Model's inputs are processed via a channel‑wise normalization layer that enforced zero mean and bounded amplitude independently for each channel. For the EEG signals, as we are using sampling frequency of 100 Hz, wavelet coefficients corresponding to the $\delta^/*$ (0–3.125 Hz), $\theta^*$ (3.125–6.25 Hz), $\alpha^*$ (6.25–12.5 Hz), $\beta^*$ (12.5–25 Hz), and $\gamma^*$ (25–50 Hz) frequency bands were extracted using a custom layer implementing a 4-level RDWT with a Symlet-2 MW. To mitigate contamination from motion-related artifacts in frequency components above 25 Hz, sparsity was explicitly enforced by masking and setting the $\gamma^*$-band coefficients to zero.
\\
The proposed wavelet-domain filtering layer operates on RDWT coefficients $X\in R^{(B\times C\times W\times (D+1))}$, filtering each sub-band independently for each EEG channel using a dedicated convolution-based zero-phase filtering scheme. This scheme is implemented using two sequential 1D convolutional operations, initialized as identity filters and made trainable, with the second convolution applied to the time-reversed signal to eliminate phase distortion. The custom layer allows kernel sizes to be specified as configurable arguments. Kernel sizes were selected to provide nearly scale-matched temporal support, with effective frequency scales inversely proportional to kernel length ($f_{eff}\simeq F_s/K$); Table~\ref{tab:Table2} summarizes the rationale behind the kernel sizes used for each sub-band. The filtered coefficients are stacked across channels and sub-bands, producing an output tensor of the same shape as the input coefficients. Finally, the denoised EEG channel signals are reconstructed using a four-level inverse RDWT layer. 
\\
Additional channel-wise temporal filtering was performed using a 2D convolution layer with a single filter and a kernel size of (1, 63), followed by a ‘tanh’ activation. Spatial filtering was subsequently performed in a depth-wise convolution manner across EEG channels using a 2D convolution layer with a single filter and a kernel size of (6, 1), followed by a ‘tanh’ activation function, yielding a single-channel EEG feature representation.
\\
Once the Model is trained, the output of the spatial filter (second 2D convolution layer) is used as the EEG feature for analysis. After the spatial filtering, a stack of causal 1D convolutional layers was applied—first with 64 filters of size 64, then 32 filters of size 32, followed by 16 filters of size 16, and finally an 8-filter layer with a kernel size of 3 and a dilation rate of 2, all using `ELU'. The filtered features were then processed by two recurrent layers: a bidirectional LSTM with 32 units and \texttt{return\_sequences}=True, followed by a unidirectional LSTM with 16 units. Finally, a fully connected layer (`ReLU' activation) was imposed to generate the output sequence - 256 samples corresponding to the predicted ocular trajectory.
\begin{figure*}[!t]
\centering
\includegraphics[width=\textwidth]{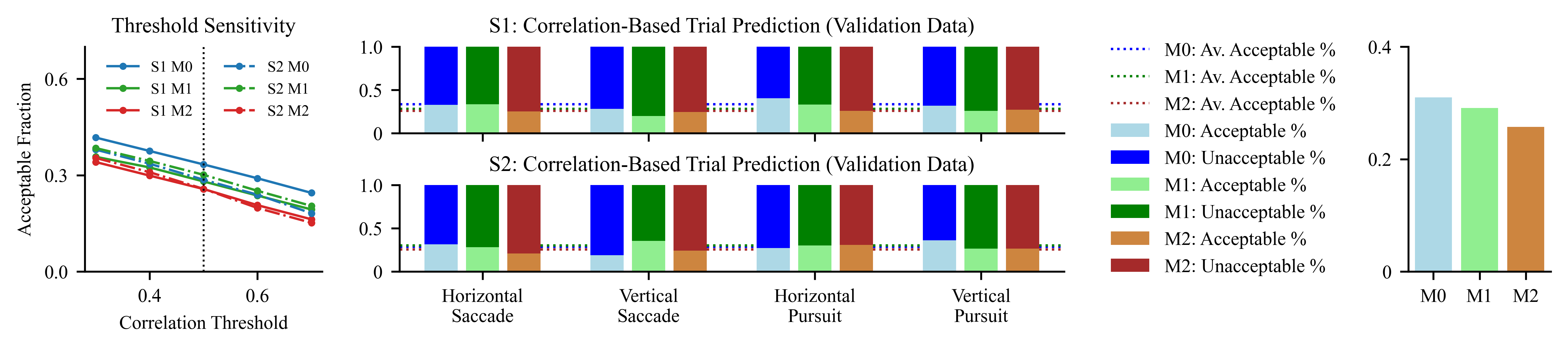}
\caption{Validation-set correlation-based outcomes for M0–M2. The leftmost panel shows mean acceptable-trial fraction versus correlation threshold (0.3–0.7) for S1 and S2, confirming the ordering M0 $>$ M1 $>$ M2 is preserved across thresholds. Stacked bars show the percentage of acceptable (r $>$ 0.5) and unacceptable (r $\le$ 0.5) trials per task for S1 and S2; dashed lines mark participant-wise mean acceptable rates. The rightmost panel shows cross-subject model averages.}
\label{fig7}
\end{figure*}
\subsubsection{Model Training \& Ablation Study}
\begin{figure}[!ht]
\vspace{-10pt}
\centering
\includegraphics[width=\linewidth]{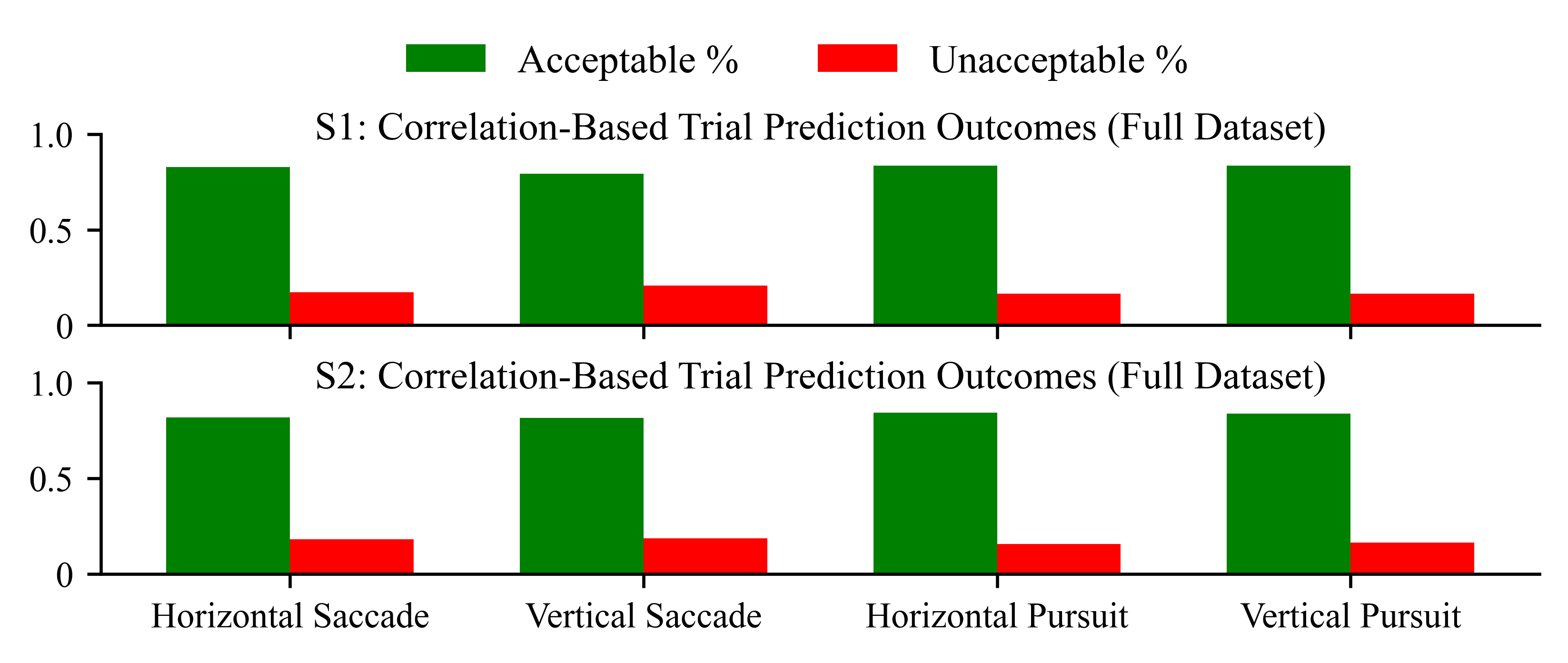}
\caption{M0-Pearson-correlation outcomes (full dataset). Green and red bars indicate valid (r $>$ 0.5) and invalid (r $\le$ 0.5) trial windows, respectively.}
\label{fig8}
\end{figure}

For each VOMS trial, a sliding-window approach was applied with a 2.56-sec window and a 200 ms stride. The EEG segment within each window served as the input feature, while the corresponding windowed target trajectory, represented via L2 distance, served as the target for model training. Data were partitioned chronologically, with the initial 80\% of samples used for training and the final 20\% for validation. Because EEG signals are temporally autocorrelated, a random split would place adjacent, highly similar windows in both partitions; a chronological split was therefore preferred to avoid leaking information and inflating validation performance. The Model was trained for up to 300 epochs using early stopping (patience=50) and a batch size of 64. Model predictions were validated against the target trajectory by computing the Pearson correlation coefficient. Trial windows were deemed acceptable (valid) when the correlation between the predicted and target trajectories was at least 0.5.
\\
To validate the RDWT-inspired filtering stage using the Symlet-2 MW, we performed an ablation study, where we used 3 models: \textbf{Model\#0 (M0)} – complete model architecture as shown in Fig.~\ref{fig5}, \textbf{Model\#1 (M1)} – A model without a wavelet-domain filtering layer, in which wavelet coefficients are not filtered using zero-phase-shift trainable convolution (excluding the light-cyan wavelet-domain filter block shown in Fig.~\ref{fig5}). \textbf{Model\#2 (M2)} – A model without RDWT-related layers (excluding the light green RDWT, masking and inverse-RDWT blocks as well as the light-cyan wavelet-domain filter block shown in Fig.~\ref{fig5}).
\\
Ablation results, as illustrated in Fig.~\ref{fig7}, indicate that the complete model (M0) consistently achieves the highest proportion of acceptable trials across tasks and subjects, while removing wavelet-domain filtering (M1) or all RDWT-related components (M2) leads to a systematic reduction in performance. These findings highlight the importance of wavelet-domain processing; therefore, the complete model (M0) was used for all subsequent analyses. Fig.~\ref{fig8} illustrates the percentage of acceptable (valid) and unacceptable (discarded) trial windows across VOMSs and subjects, using the full dataset evaluated with the M0 model.
\subsubsection{Feature Extraction}
\begin{figure}[!hb]
\centering
\includegraphics[width=\linewidth]{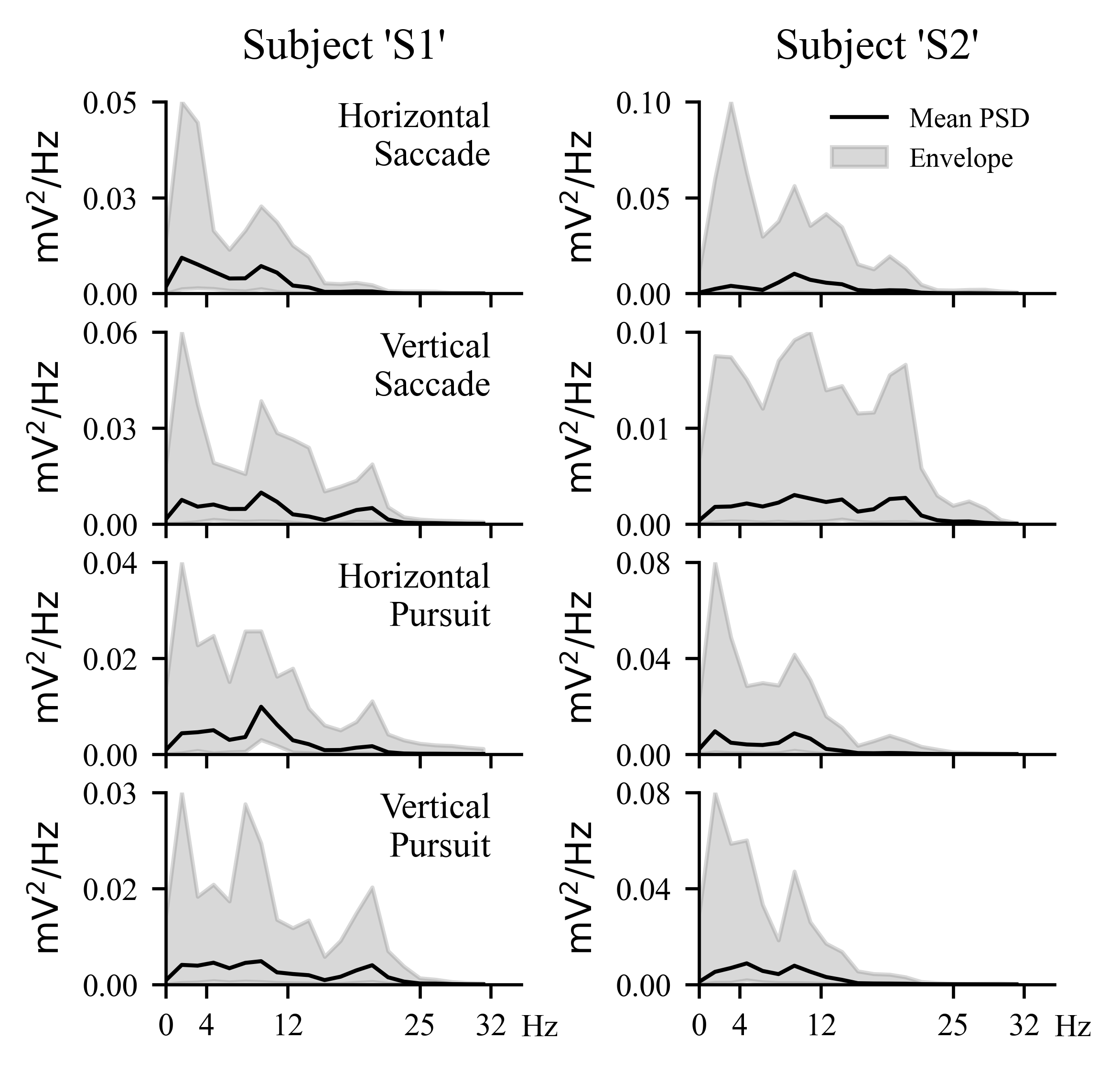}
\caption{PSD plots of extracted EEG features for valid windows, showing mean and min–max envelopes across VOMSs and subjects. Rows (top to bottom): Horizontal Saccade, Vertical Saccade, Horizontal Pursuit, and Vertical Pursuit.}
\label{fig9}
\end{figure}

For all valid trial windows identified using the M0 (complete architecture) model,  features were extracted from the network after spatial filtering across channels, specifically from the output of the second 2D convolution layer following the tanh activation. This corresponds to the dark gray–shaded block shown in Fig.~\ref{fig5} and yields a single-channel temporal feature representation for each trial window. These features capture the dominant, spectrally and spatially informed temporal dynamics of EEG activity that are most relevant for downstream temporal modeling and prediction.
\\
We evaluated the power spectral density (PSD) of the extracted EEG features for all valid windows, plotting the mean PSD together with a min–max envelope to illustrate inter-window dispersion across VOMSs and subjects, as shown in Fig.~\ref{fig9}.
\subsubsection{Ocular Response Time Analysis}
\begin{figure}[!ht]
\centering
\includegraphics[width=\linewidth]{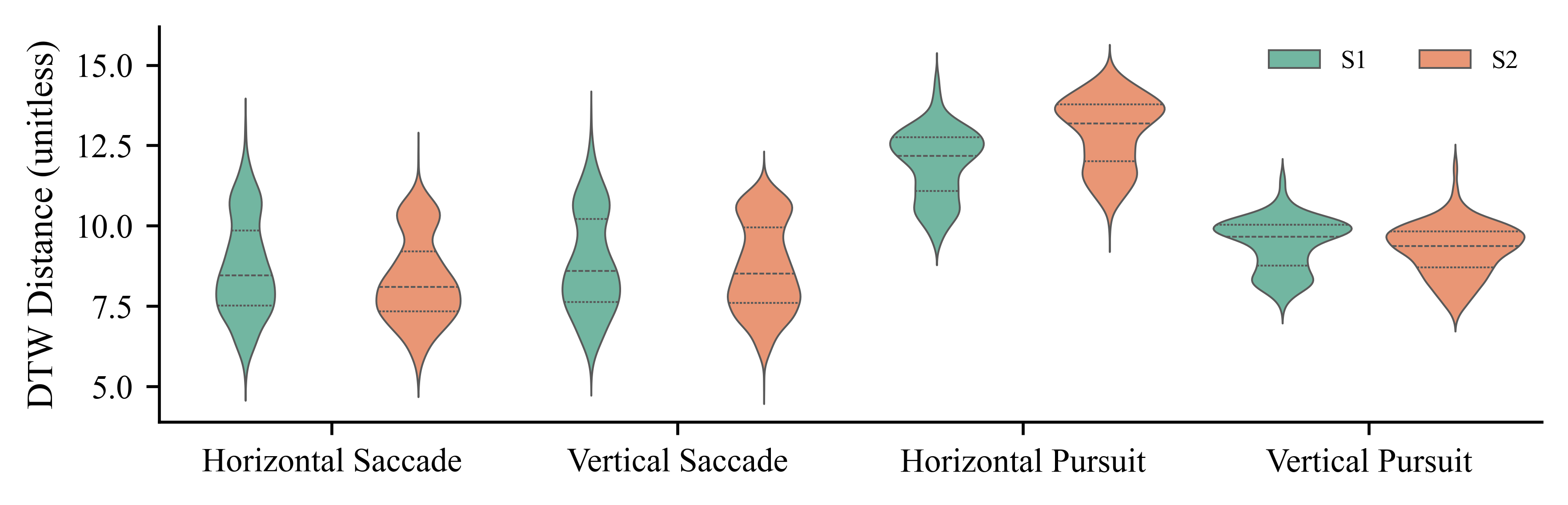}
\caption{Violin plots showing DTW distance distributions between extracted EEG features and target trajectories across VOMSs and subjects. Horizontal lines indicate the median and interquartile range (IQR).}
\label{fig10}
\vspace{-10pt}
\end{figure}

The Dynamic Time Warping (DTW) distance is commonly used to analyze temporal delays and alignment differences between two time series, especially when they have similar shapes but are misaligned in time \cite{Giorgino2009}. To quantify ocular response time within the valid windows, the DTW distance was computed between the extracted EEG features from the trained M0 model (after depth-wise convolution) and the target trajectory, using a constrained alignment window of 50 samples ($\pm$500 ms at 100 Hz) to limit excessive temporal warping. A DTW window of 50 samples was chosen to capture the typical neural responses associated with fixations and the preparation of subsequent eye movements, which predominantly occur within $\sim$300 ms \cite{Nikolaev2016}. A Violin Plot (Fig.~\ref{fig10}) was used to visualize the DTW distance across VOMSs and subjects, with horizontal lines indicating the median and interquartile range (IQR), clearly showing similar patterns across subjects for the respective VOMS, with shifted medians and IQRs. 
\\
Furthermore, we performed a nonparametric Mann–Whitney U test to assess window-level differences in DTW distance between S1 and S2 for each VOMS task, using a significance threshold of $\alpha$ = 0.05. Additionally, for valid windows, we plotted the normalized cross-correlation dynamics across different VOMSs for both participants (Fig.~\ref{fig11}), using DTW-aligned extracted features and target trajectories. Fig.~\ref{fig11} reveals temporal alignment and response dynamics across subjects and VOMSs, while the envelopes help understand inter-window variability and provide insight into consistency. 
\section{Results}
As illustrated in Fig.~\ref{fig4}, applying a 4th order Butterworth band-pass filter (0.5-30 Hz) followed by average referencing effectively mitigated DC drift and outliers in the raw EEG. 
When aggregated across all VOMSs and subjects, the filtered data achieved an overall mean of $-2.91 \times 10^{-8}\,\mathrm{mV}$ with a standard deviation of $0.009\,\mathrm{mV}$.
\\
Fig.~\ref{fig9} illustrates the results of the PSD analysis of valid EEG feature windows, highlighting distinct and task-specific spectral characteristics. The PSD for horizontal saccades is dominated by low-frequency components, with most power concentrated below $\sim10$ Hz. Vertical saccades also show strong low-frequency dominance, but with a more pronounced contribution in the $\beta$-band compared to horizontal saccades. The PSD during horizontal and vertical pursuit exhibits a broader spectral distribution, with elevated power extending from low frequencies into the $\alpha$ band, and additional $\beta$-band contributions observed particularly for participant S1.
\\
Subject-wise comparison of DTW distances across VOMSs (Fig.~\ref{fig10}) shows differences between the two participants and a task-dependent pattern, suggesting that task-specific ocular response time—quantified through DTW—is a candidate qualitative metric that may support distinguishing subjects with mTBI in future work. For both participants, Fig.~\ref{fig10} reveals that saccadic tasks exhibit lower median DTW distances; however, the distributions span a broader range, reflecting trial-to-trial variability despite a consistent central tendency. In contrast, horizontal and vertical pursuit tasks demonstrate higher median DTW distances with comparatively narrower distributions, indicating more consistent but temporally delayed ocular response timing relative to saccadic tasks.
\\
At the window level, a Mann–Whitney U test showed DTW distance distributions differed between participants (S1, S2) across all VOMS tasks (Table~\ref{tab:Table3}). A sensitivity analysis across thresholds 0.3–0.7 (step 0.1) preserved the per-task direction, confirming the outcomes are not a 0.5-cutoff artifact. As these are repeated measures from two individuals, p-values are reported descriptively, not as subject-level inference. The per-task separation nonetheless supports DTW-derived ocular response time as a candidate metric for future mTBI assessment, since mTBI is associated with elevated response times \cite{Churchill2021, Christensen2023, Mullen2014} and poorer timing consistency in pursuit tasks \cite{Stubbs2019}. Group-level testing is deferred to a larger, mTBI-inclusive cohort.
\begin{table}[h]
\centering
\caption{Window-level DTW differences between participants (S1, S2) across VOMS tasks (Mann-Whitney U test).}
\label{tab:Table3}
\renewcommand{\arraystretch}{1.3}
\setlength{\tabcolsep}{4pt}
\begin{tabular}{lcccc}
\hline
\textbf{VOMS} & \textbf{P-value} & \textbf{S1 Median} & \textbf{S2 Median} & \textbf{Direction} \\
\hline
Horizontal Saccade & $<$0.0001 & 8.454  & 8.104  & S1 $>$ S2 \\
Vertical Saccade   & $<$0.0001 & 8.599  & 8.497  & S1 $>$ S2 \\
Horizontal Pursuit & $<$0.0001 & 12.175 & 13.174 & S1 $<$ S2 \\
Vertical Pursuit   & $<$0.0001 & 9.663  & 9.371  & S1 $>$ S2 \\
\hline
\multicolumn{5}{l}{\scriptsize P-values are window-level across 2 participants, not subject-level inference.} \\
\end{tabular}
\end{table}
\begin{figure}[!b]
\centering
\includegraphics[width=0.9\linewidth]{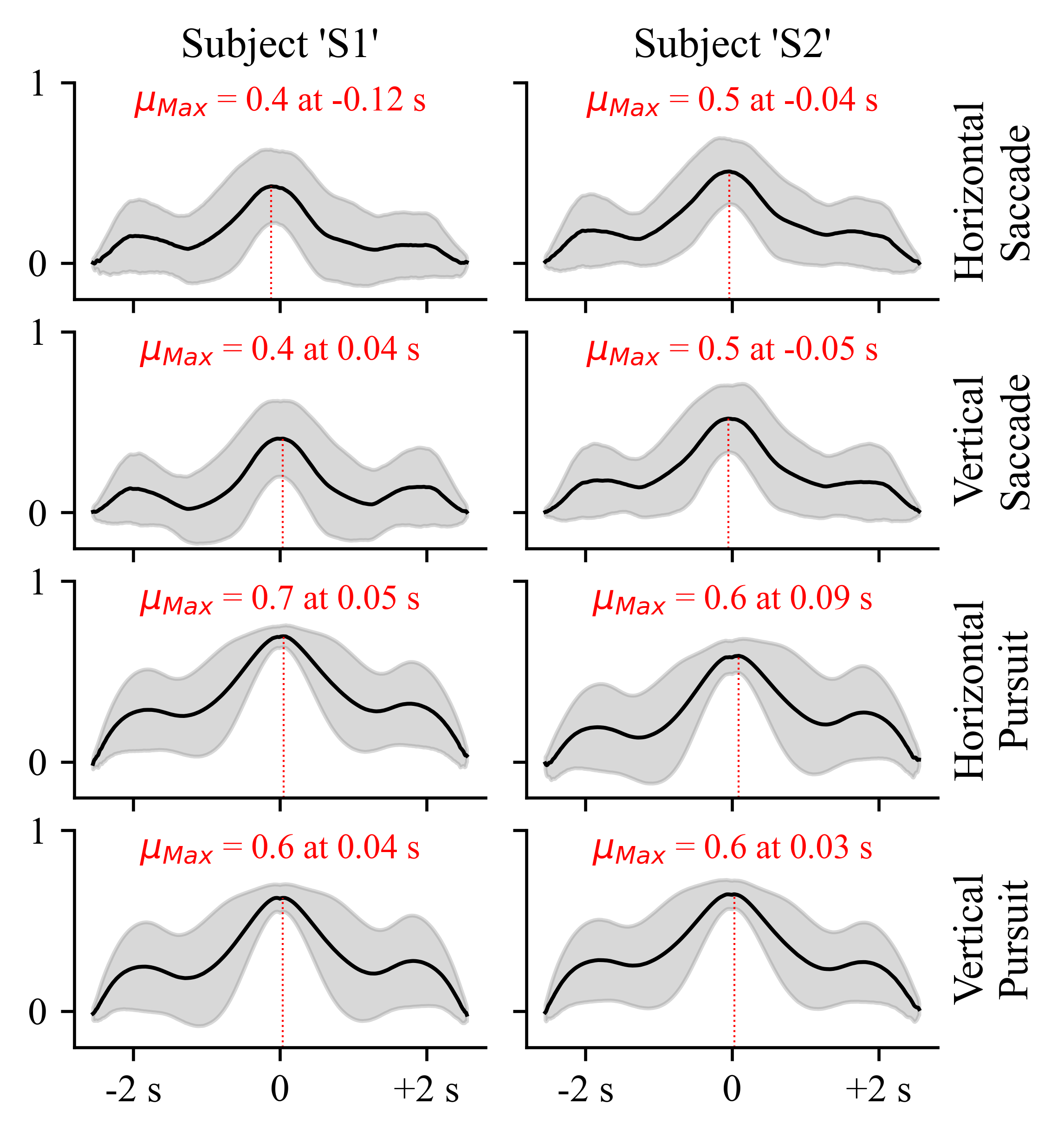}
\caption{Normalized cross-correlation dynamics (unitless) between DTW-aligned EEG features and target trajectories across VOMSs and subjects, computed over valid windows (r $>$ 0.5).}
\label{fig11}
\end{figure}
\\
Fig.~\ref{fig11} shows the normalized cross-correlation dynamics across VOMSs for both subjects, using the DTW-aligned extracted feature and the target trajectory (L2 distance). The lag or lead at the point of maximum cross-correlation (uMax) provides a temporal error metric that indicates residual misalignment after DTW alignment. For the horizontal saccade task, S1 exhibits a pronounced lead exceeding 100 ms, whereas for the remaining VOMSs the lag/lead remains below 100 ms.  For pursuit tasks, the DTW-aligned feature consistently lags the target VOMS trajectory, reflecting reactive tracking of the AR patch; in contrast, for saccade tasks, it leads the target (except S1 vertical saccade), indicating anticipatory positioning. Additionally, pursuit tasks also yield the highest peak correlations (0.6–0.7), reflecting stronger temporal coupling between the aligned feature and the target trajectory. Taken together, these properties make horizontal and vertical pursuits the most suitable VOMSs.
\section{Conclusion}
This study demonstrates that EEG signals preprocessed using BPF and average referencing, followed by feature extraction via an RDWT-driven deep neural framework, can effectively support downstream analysis of subject- and VOMS-specific ocular response time using DTW. Spectral characterization via PSD analysis revealed task-specific and consistent spectral profiles, with dominant power concentrated in characteristic frequency ranges across trials.
DTW-derived ocular response time emerged as a promising qualitative metric, showing task-dependent, window-level differences between participants that motivate group-level testing in a larger cohort. Cross-correlation analysis further highlighted distinct temporal strategies, with pursuit tasks exhibiting reactive tracking (positive lag) and strong temporal coupling. Taken together, these findings show that RDWT-based EEG feature extraction combined with DTW-distance-based temporal alignment provides a reliable and informative framework for capturing temporal fidelity—especially within pursuit paradigms—distinguishing subject groups and motivating future mTBI-related assessments.
\\
Future work will focus on integrating EOG tracking and real-time removal of ocular artifacts from EEG using adaptive filtering approaches such as H-infinity [27], alongside real-time eye tracking derived directly from EOG signals. We also plan to incorporate simultaneous EEG and real-time eye-tracking trajectories, together with the target stimulus trajectory, to enable a more comprehensive characterization of neuro-oculo-temporal dynamics. In addition, upcoming experiments will involve a larger and more diverse participant cohort, including individuals with mTBI, to strengthen inter-subject comparisons and enhance the generalizability of our findings.
\section*{Acknowledgment}
We gratefully acknowledge Dr.\ Jerome Jeevarajan, MD (Department of Neurology, McGovern Medical School), for reviewing this work and providing clinically informed feedback that helped strengthen its clinical relevance. We also acknowledge NSF REU (2024) student Nathaniel Liu (University of Arkansas, Biomedical Engineering) for his contributions to data collection and analysis during the initial phase of the project. In addition, we thank undergraduate student Brandon Burlison (University of Houston, Industrial Design) for his assistance with data collection.





\begin{thebibliography}{99}
\bibitem{CDC2025} CDC, ``Traumatic Brain Injury \& Concussion: Facts about TBI,'' 2025. [Online]. Available: \url{https://www.cdc.gov/traumatic-brain-injury/data-research/facts-stats/index.html}. Accessed:  Feb. 20, 2026.

\bibitem{DoD_TBI_No} DoD, ``DoD TBI Worldwide Numbers,'' Military Health System, 2026. [Online]. Available: \url{https://www.health.mil/Military-Health-Topics/Centers-of-Excellence/Traumatic-Brain-Injury-Center-of-Excellence/DOD-TBI-Worldwide-Numbers}. Accessed: Feb. 20, 2026.

\bibitem{McKee2014} A. C. McKee and M. E. Robinson, ``Military-related traumatic brain injury and neurodegeneration,'' \textit{Alzheimer's \& Dementia: J. Alzheimer's Assoc.}, vol. 10, no. 3, Suppl., pp. S242--S253, 2014. [Online]. Available: \url{https://doi.org/10.1016/j.jalz.2014.04.003}.

\bibitem{ImPACT} T. Covassin, R. J. Elbin, J. L. Stiller-Ostrowski, and A. P. Kontos, ``Immediate post-concussion assessment and cognitive testing (ImPACT) practices of sports medicine professionals,'' \textit{J. Athletic Training}, vol. 44, no. 6, pp. 639--644, 2009. [Online]. Available: \url{https://doi.org/10.4085/1062-6050-44.6.639}.

\bibitem{PCSS} P. Langevin, P. Fr\'emont, P. Fait, and J.-S. Roy, ``Responsiveness of the post-concussion symptom scale to monitor clinical recovery after concussion or mild traumatic brain injury,'' \textit{Orthopaedic J. Sports Med.}, vol. 10, no. 10, p. 23259671221127049, 2022. [Online]. Available: \url{https://doi.org/10.1177/23259671221127049}.

\bibitem{BESS} D. R. Bell, K. M. Guskiewicz, M. A. Clark, and D. A. Padua, ``Systematic review of the balance error scoring system,'' \textit{Sports Health}, vol. 3, no. 3, pp. 287--295, 2011. [Online]. Available: \url{https://doi.org/10.1177/1941738111403122}.

\bibitem{Farnsworth2017} J. L. Farnsworth, L. Dargo, B. G. Ragan, and M. Kang, ``Reliability of computerized neurocognitive tests for concussion assessment: A meta-analysis,'' \textit{J. Athletic Training}, vol. 52, no. 9, pp. 826--833, 2017. [Online]. Available: \url{https://doi.org/10.4085/1062-6050-52.6.03}.

\bibitem{Resch2013} J. Resch, A. Driscoll, N. McCaffrey, C. Brown, M. S. Ferrara, S. Macciocchi, T. Baumgartner, and K. Walpert, ``ImPACT test--retest reliability: reliably unreliable?'' \textit{J. Athletic Training}, vol. 48, no. 4, pp. 506--511, 2013. [Online]. Available: \url{https://doi.org/10.4085/1062-6050-48.3.09}.

\bibitem{Mason2020} S. J. Mason, B. S. Davidson, M. Lehto, A. Ledreux, A.-C. Granholm, and K. A. Gorgens, ``A cohort study of the temporal stability of ImPACT scores among NCAA Division I collegiate athletes: Clinical implications of test-retest reliability for enhancing student athlete safety,'' \textit{Arch. Clin. Neuropsychol.}, vol. 35, no. 7, pp. 1131--1144, 2020. [Online]. Available: \url{https://doi.org/10.1093/arclin/acaa047}.

\bibitem{Uriguen2015} J. A. Urig\"uen and B. Garcia-Zapirain, ``EEG artifact removal: State-of-the-art and guidelines,'' \textit{J. Neural Eng.}, vol. 12, no. 3, p. 031001, 2015. [Online]. Available: \url{https://doi.org/10.1088/1741-2560/12/3/031001}.

\bibitem{Craik2023} A. Craik, J. J. Gonz\'alez-Espa\~na, A. Alamir, D. Edquilang, S. Wong, L. S\'anchez Rodr\'iguez, J. Feng, G. E. Francisco, and J. L. Contreras-Vidal, ``Design and validation of a low-cost mobile EEG-based brain--computer interface,'' \textit{Sensors}, vol. 23, no. 13, p. 5930, 2023. [Online]. Available: \url{https://doi.org/10.3390/s23135930}.

\bibitem{Kutcher2014} J. S. Kutcher and C. C. Giza, ``Sports concussion diagnosis and management,'' \textit{Continuum (Minneapolis, Minn.)}, vol. 20, no. 6, Sports Neurology, pp. 1552--1569, 2014. [Online]. Available: \url{https://doi.org/10.1212/01.CON.0000458974.78766.58}.

\bibitem{Munia2017} T. T. K. Munia, A. Haider, C. Schneider \textit{et al.}, ``A novel EEG based spectral analysis of persistent brain function alteration in athletes with concussion history,'' \textit{Sci. Rep.}, vol. 7, p. 17221, 2017. [Online]. Available: \url{https://doi.org/10.1038/s41598-017-17414-x}.

\bibitem{Yue2020} J. K. Yue, P. S. Upadhyayula, L. N. Avalos, H. Deng, and K. K. W. Wang, ``The role of blood biomarkers for magnetic resonance imaging diagnosis of traumatic brain injury,'' \textit{Medicina}, vol. 56, no. 2, p. 87, 2020. [Online]. Available: \url{https://doi.org/10.3390/medicina56020087}.

\bibitem{Mucha2014} A. Mucha, M. W. Collins, R. J. Elbin, J. M. Furman, C. Troutman-Enseki, R. M. DeWolf, G. Marchetti, and A. P. Kontos, ``A brief Vestibular/Ocular Motor Screening (VOMS) assessment to evaluate concussions: Preliminary findings,'' \textit{Amer. J. Sports Med.}, vol. 42, no. 10, pp. 2479--2486, 2014. [Online]. Available: \url{https://doi.org/10.1177/0363546514543775}.

\bibitem{Toda1993} K. Toda, H. Tachibana, M. Sugita, and K. Konishi, ``P300 and reaction time in Parkinson's disease,'' \textit{J. Geriatr. Psychiatry Neurol.}, vol. 6, no. 3, pp. 131--136, 1993. [Online]. Available: \url{https://doi.org/10.1177/089198879300600301}.

\bibitem{vanderGeest2001} J. N. van der Geest, C. Kemner, G. Camfferman, M. N. Verbaten, and H. van Engeland, ``Eye movements, visual attention, and autism: A saccadic reaction time study using the gap and overlap paradigm,'' \textit{Biol. Psychiatry}, vol. 50, no. 8, pp. 614--619, 2001. [Online]. Available: \url{https://doi.org/10.1016/s0006-3223(01)01070-8}.

\bibitem{Wylie2009} S. A. Wylie, W. P. van den Wildenberg, K. R. Ridderinkhof, T. R. Bashore, V. D. Powell, C. A. Manning, and G. F. Wooten, ``The effect of Parkinson's disease on interference control during action selection,'' \textit{Neuropsychologia}, vol. 47, no. 1, pp. 145--157, 2009. [Online]. Available: \url{https://doi.org/10.1016/j.neuropsychologia.2008.08.001}.

\bibitem{Zhang2016} J. Zhang, T. Rittman, C. Nombela, A. Fois, I. Coyle-Gilchrist, R. A. Barker, L. E. Hughes, and J. B. Rowe, ``Different decision deficits impair response inhibition in progressive supranuclear palsy and Parkinson's disease,'' \textit{Brain}, vol. 139, no. 1, pp. 161--173, 2016. [Online]. Available: \url{https://doi.org/10.1093/brain/awv331}.

\bibitem{Xie2024} X. Xie, R. Zhou, Z. Fang, Y. Zhang, Q. Wang, and X. Liu, ``Seeing beyond words: Visualizing autism spectrum disorder biomarker insights,'' \textit{Heliyon}, vol. 10, no. 9, p. e30420, 2024. [Online]. Available: \url{https://doi.org/10.1016/j.heliyon.2024.e30420}.

\bibitem{Pope2022} K. J. Pope, T. W. Lewis, S. P. Fitzgibbon, A. S. Janani, T. S. Grummett, P. A. H. Williams, M. Battersby, T. Bastiampillai, E. M. Whitham, and J. O. Willoughby, ``Managing electromyogram contamination in scalp recordings: An approach identifying reliable beta and gamma EEG features of psychoses or other disorders,'' \textit{Brain Behav.}, vol. 12, no. 9, p. e2721, 2022. [Online]. Available: \url{https://doi.org/10.1002/brb3.2721}.

\bibitem{Sarkar2025} S. Sarkar and J. L. Contreras-Vidal, ``Optimal discrete mother wavelet selection for EEG motor imagery decoding: A comparative study,'' in \textit{Health Informatics and Medical Systems and Biomedical Engineering}, ser. Communications in Computer and Information Science, vol. 2259. Cham: Springer, 2025. [Online]. Available: \url{https://doi.org/10.1007/978-3-031-85908-3_2}.

\bibitem{Giorgino2009} T. Giorgino, ``Computing and visualizing dynamic time warping alignments in R: The dtw package,'' \textit{J. Stat. Softw.}, vol. 31, no. 7, pp. 1--24, 2009. [Online]. Available: \url{https://doi.org/10.18637/jss.v031.i07}.

\bibitem{Nikolaev2016} A. R. Nikolaev, R. N. Meghanathan, and C. van Leeuwen, ``Combining EEG and eye movement recording in free viewing: Pitfalls and possibilities,'' \textit{Brain Cogn.}, vol. 107, pp. 55--83, 2016. [Online]. Available: \url{https://doi.org/10.1016/j.bandc.2016.06.004}.


\bibitem{Churchill2021} N. W. Churchill, M. G. Hutchison, S. J. Graham, and T. A. Schweizer, ``Brain function associated with reaction time after sport-related concussion,'' \textit{Brain Imag. Behav.}, vol. 15, no. 3, pp. 1508--1517, 2021. [Online]. Available: \url{https://doi.org/10.1007/s11682-020-00349-9}.


\bibitem{Christensen2023} B. A. Christensen, B. Clark, A. M. Muir, W. D. Allen, E. M. Corbin, T. Jaggi, N. Alder, A. Clawson, T. J. Farrer, E. D. Bigler, and M. J. Larson, ``Interhemispheric transfer time and concussion in adolescents: A longitudinal study using response time and event-related potential measures,'' \textit{Front. Hum. Neurosci.}, vol. 17, p. 1161156, 2023. [Online]. Available: \url{https://doi.org/10.3389/fnhum.2023.1161156}.

\bibitem{Mullen2014} S. J. Mullen, Y. H. Y\"ucel, M. Cusimano, T. A. Schweizer, A. Oentoro, and N. Gupta, ``Saccadic eye movements in mild traumatic brain injury: A pilot study,'' \textit{Can. J. Neurol. Sci.}, vol. 41, no. 1, pp. 58--65, 2014. [Online]. Available: \url{https://doi.org/10.1017/S0317167100016279}.

\bibitem{Stubbs2019} J. L. Stubbs, S. L. Corrow, B. R. Kiang, J. C. Corrow, H. L. Pearce, A. Y. Cheng, J. J. S. Barton, and W. J. Panenka, ``Working memory load improves diagnostic performance of smooth pursuit eye movement in mild traumatic brain injury patients with protracted recovery,'' \textit{Sci. Rep.}, vol. 9, no. 1, p. 291, 2019. [Online]. Available: \url{https://doi.org/10.1038/s41598-018-36286-3}.


\end{thebibliography}
\end{document}